\documentclass[final,10pt,twocolumn,numbers]{elsarticle}
\biboptions{sort&compress}

\usepackage{amssymb}
\usepackage{amsmath}

\begin{document}

\begin{frontmatter}



\title{Low-Field J-Coupling Spectroscopy Measured with a Diamond Nitrogen Vacancy Spectrometer} 

\author[ireap,qtc]{Stephen J. DeVience\corref{cor1}}
\ead{devience@umd.edu}
\cortext[cor1]{Corresponding Author}
\author[qtc]{John W. Blanchard}
\author[chem]{Camille G. McDonnell}
\author[qtc,phys,elec]{Ronald L. Walsworth}
\ead{walsworth@umd.edu}
\affiliation[ireap]{organization={Institute for Research in Electronics and Applied Physics, University of Maryland},
             addressline={8279 Paint Branch Drive},
             city={College Park},
             postcode={20742},
             state={MD},
             country={USA}}
\affiliation[qtc]{organization={Quantum Technology Center, University of Maryland},
             addressline={2218 Kim Engineering Building},
             city={College Park},
             postcode={20742},
             state={MD},
             country={USA}}
\affiliation[chem]{organization={Department of Chemical Physics, University of Maryland},
            addressline={8108 Regents Drive}, 
            city={College Park},
            postcode={20742}, 
            state={MD},
            country={USA}}
\affiliation[phys]{organization={Department of Physics, University of Maryland},
            addressline={4150 Campus Drive}, 
            city={College Park},
            postcode={20742}, 
            state={MD},
            country={USA}}
\affiliation[elec]{organization={Department of Electrical and Computer Engineering, University of Maryland},
            addressline={8223 Paint Branch Drive}, 
            city={College Park},
            postcode={20742}, 
            state={MD},
            country={USA}}

\begin{abstract}
Nitrogen vacancy centers in diamond (NVs) are point defects well suited to quantum sensing of magnetic fields. The NV spin states can be initialized and read out optically with high spatial resolution, enabling detection of picoliter-scale volumes of nuclear spins. Using Coherently Averaged Synchronized Readout (CASR), standard NMR spectroscopy techniques such as free decay, spin echo, and PGSE have been performed with Hz-scale spectral resolution using NV-diamond sensors. However, using NV detection for conventional high-resolution NMR at multi-tesla fields is challenging due to the need to manipulate NV electronic spins at resonance frequencies 100 GHz or higher. Low-field J-coupling spectroscopy provides an alternative path to chemical resolution in a regime more easily accessible to NVs.

Here, we demonstrate homonuclear and heteronuclear J-coupling spectroscopy using an NV-NMR spectrometer based on permanent magnets operating at $\sim$890 kHz proton Larmor frequency (i.e., bias magnetic field $\sim$21 mT). This choice of field strength enables chemical characterization based on J-coupling using sequences such as Synchronized Echo (SyncE) and Spin-Echo Double Resonance (SEDOR), which do not require the strong chemical shift dispersion and pristine field homogeneity of high-field NMR. The frequency range is easily detected by NVs using standard XY dynamical decoupling sequences and is well-suited for applying Overhauser DNP, which can boost proton signal by two orders of magnitude.

We acquire initial NV-NMR measurements of T$_1$, T$_2$, and Overhauser DNP in glycerol before advancing to SEDOR spectroscopy of methanol-$^{13}$C and a methanol-$^{13}$C/acetonitrile-2-$^{13}$C mixture at sample volumes $\sim$50 pL, in which we demonstrate measurement of $^1J_\text{CH}$ couplings and resolution between the species. We also acquire a homonuclear J-coupling spectrum of a natural-abundance propylene carbonate sample.
\end{abstract}






\end{frontmatter}



\section{Introduction}
\label{intro}

Nitrogen vacancy centers in diamond (NVs) are optically active point defects that are well suited to quantum sensing of magnetic fields from small samples \cite{Levine2018}. NV electronic spin states can be optically initialized and read out via photoluminescence (PL) with high spatial resolution, enabling the detection of small volumes of nuclear spins from the nanoscale to microscale. Early work used dynamical decoupling sequences applied to electronic spins to perform AC magnetometry on the magnetic noise created by small numbers of nuclear spins (i.e., statistical polarization), allowing different nuclear species such as $^1$H and $^{19}$F to be distinguished \cite{Loretz2014, Staudacher2013, Mamin2013, DeVience2015, Rugar2015, Kehayias2017}. However, the spectral resolution of these techniques was limited by the $\sim$5 ms T$_1$ electronic spin relaxation time of the NV center. The subsequent development of extended readout techniques such as Coherently Averaged Synchronized Readout (CASR) \cite{glenn_2018} and Quantum Heterodyne Detection (QDyne) \cite{qdyne} removed this limit and enabled {NV-NMR} with standard spectroscopy techniques, such as free decay, spin echo, and {PGSE}, of both thermally polarized and hyperpolarized samples \cite{Bruckmaier2023, Bucher2019, Bucher_2020}. The spectral resolution of CASR is not limited by NV relaxation times, and NV-NMR linewidths below 1 Hz have been achieved in systems with high field homogeneity \cite{Smits2019}.

Despite these advances, NV-NMR has been primarily performed at relatively low field strengths (<500 mT) where chemical shift resolution is poor, with demonstrations mostly confined to relaxometry \cite{Bruckmaier2023} and to spectroscopy of carefully selected molecules either with large chemical shifts \cite{glenn_2018, Bucher_2020} or with large splittings induced by heteronuclear J-coupling \cite{Smits2026, Smits2019, Kehayias2017}. Using NV-NMR for general chemical identification and structure determination would require increasing the field strength by roughly an order of magnitude, which is challenging for two reasons. First, the NV electronic spin states must be controlled with microwave pulses at a frequency proportional to 28 GHz/T. In typical high-field NMR regimes of many teslas, the synthesis, amplification, and delivery of the required pulses become difficult. Second, the inter-pulse spacing for NV dynamical decoupling sequences used for NMR signal detection is inversely proportional to the NMR Larmor frequency and becomes impractically short for signals above a few MHz. Although alternative NV-NMR methods such as AERIS \cite{munuera-javaloy_2023} and DRACAERIS \cite{daly} have been proposed, which detect driven longitudinal oscillations of the nuclear spins rather than Larmor precession, these require uncommon cuts of diamond and have not yet been experimentally validated.

In this work, we demonstrate an alternative approach: perform NV-NMR at a low field, which is well matched to NV characteristics, and create chemical specificity using J-couplings rather than chemical shift. We first demonstrate basic functionality of an NV-NMR spectrometer operating at $\sim$21 mT, including Overhauser DNP, T$_1$, and T$_2$ measurements of a glycerol sample detected with the NV-diamond sensor. We then use NV-NMR to perform both heteronuclear and homonuclear J-coupling spectroscopy of organic molecules. For the heteronuclear spin systems methanol-$^{13}$C and acetonitrile-2-$^{13}$C, we demonstrate spin-echo double resonance (SEDOR) and its multi-echo analog that enable chemical discrimination via measurement of heteronuclear J-couplings \cite{Donaldson}. Using natural abundance propylene glycol, we perform homonuclear J-coupling spectroscopy with the synchronized echo (SyncE) sequence, which creates unique spectra for many organic molecules without the presence of a heteronucleus \cite{synce}. Our results suggest that NV-NMR and MRI of metabolites and biomolecules in small samples might be possible without high-field magnets and difficult microwave engineering.

\section{Methods}
\label{methods}
\subsection{NV-Diamond NMR Spectrometer}
\begin{figure*}[h!]
\centering
\includegraphics[width=16cm]{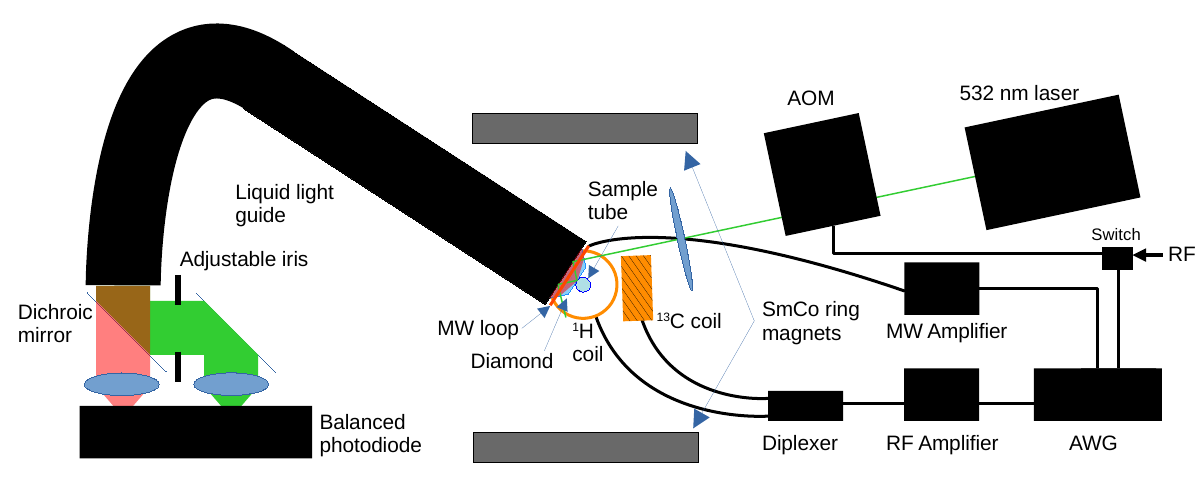}
\caption{Schematic of the low-field NV-NMR spectrometer viewed from the top. Components are described in the main text and are not drawn to scale. Detection of NMR signals from liquid in the sample tube is performed by the adjacent NV-diamond sensor via CASR NMR measurements. Pulses are first applied with orthogonal $^1$H and $^{13}$C coils to induce NMR signals from the liquid sample, held adjacent to the NV-NMR sensor. CASR NMR measurements are performed by the NVs, in which the NV electronic spins are initialized and read out with 532 nm laser pulses and manipulated by microwave (MW) pulses from the MW loop. The same MW loop is used for Overhauser DNP of the sample. Red NV photoluminescence (PL) is detected via a balanced photodiode. The static magnetic bias field (21 mT) is applied with samarium cobalt ring magnets mounted on a rotating base that enables field alignment.}\label{figsetup}
\end{figure*}

The NV-diamond NMR spectrometer (Fig.\ \ref{figsetup}) uses a 5x5x0.5 mm diamond sample with a 10 $\mu$m thick layer of NV centers at the top surface (Element Six, 2.7 ppm NV density). The NV layer is isotopically engineered with 99.999\% $^{12}$C and 17 ppm $^{15}$N. The diamond is attached to a liquid light guide (Thorlabs) that collects red NV photoluminescence (PL) and sends the light to the detection system. To increase PL collection, the diamond is cut with angled sides that reflect light emitted laterally back into the light guide. The diamond is mounted with the NV layer oriented vertically in the lab frame. For initial experiments, the liquid sample is held in a small ``cup'' against the diamond surface. In later experiments, the liquid sample is held in a thin capillary tube with 10 $\mu$m thick walls (Hampton Research) placed against the diamond surface. Green laser light (750 mW @532 nm) is directed into one side face of the diamond so that it undergoes total internal reflection and creates a 20 $\mu$m diameter spot at the point where the capillary tube touches the diamond. The green light is controlled by an acousto-optic modulator (G\&H 3250-220) to create the appropriate optical pulses for NV spin-state initialization and readout.

This NV-diamond setup sits between a pair of samarium cobalt ring magnets that create a 21 mT bias magnetic field at the center ($\sim$890 kHz $^1$H NMR frequency). The bias field is aligned with one of the four NV axes in the diamond at a 35.3$^\circ$ angle from the diamond surface; this selected class of NVs is used for NMR signal detection. RF pulses are transmitted to the NMR sample by orthogonal solenoid coils tuned to $^1$H and $^{13}$C Larmor frequencies and fed from a single source via a homebuilt diplexer. Microwave pulses are applied to the NV electronic spins by a small wire loop surrounding three sides of the diamond. All pulses are synthesized with an arbitrary waveform generator (Tektronix AWG model 7122C) and amplified to 30 W for RF and 25 W for microwaves (Mini-Circuits models LZY-22+ and ZHL-25W-63). Green (NV excitation) and red (NV PL) light collected by the liquid light guide are split into two channels of a balanced photodiode (Thorlabs PDB210A) using a dichroic mirror. The difference signal from the photodiode is collected by a TI DAQ for analysis. A piezo-controlled adjustable iris is used to control detector balance (Thorlabs ELL15).

\subsection{CASR Detection}

\begin{figure}[h!]
\centering
\includegraphics[width=8cm]{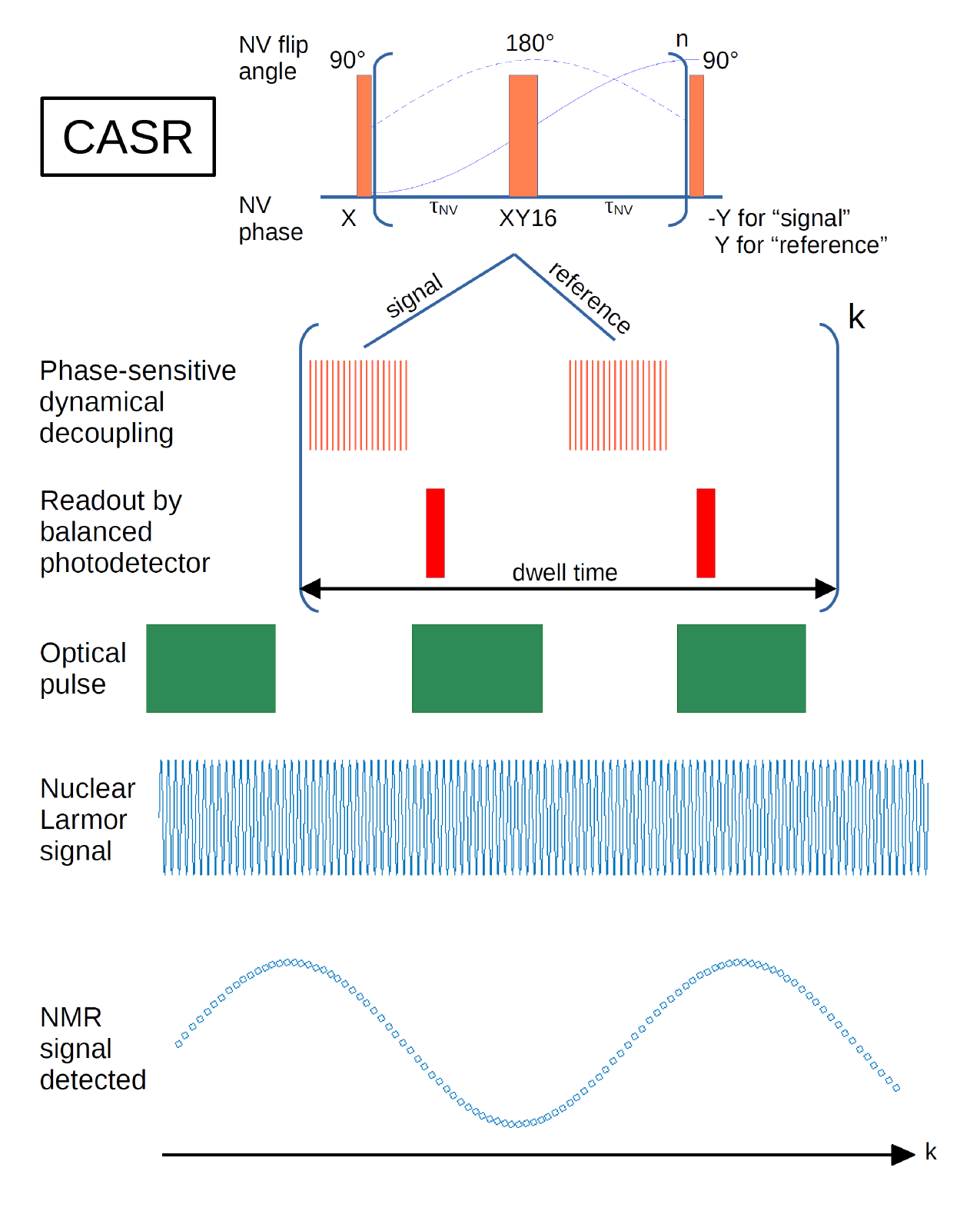}
\caption{CASR pulse sequence for high-resolution NV-NMR. A series of k phase-sensitive XY16-2 dynamical decoupling (DD) measurements are performed by the ensemble NV sensor. Each measurement consists of two blocks of NV spin-state optical initialization, an NV DD pulse sequence controlled by microwaves, and NV photoluminescence (PL) readout. The PL signals from these two blocks, called ``signal'' and ``reference,'' are subtracted to give the optical contrast for a single time point during Larmor precession. The contrast in turn depends on the phase of the NMR signal at each time point. Since the sampling rate is much smaller than the nuclear spin Larmor frequency, the resulting measurement for a series of time points is highly aliased and produces a downconverted NMR signal in a narrow spectral window.}\label{figcasr}
\end{figure}

To acquire NMR spectra using the NV-diamond sensor, optically-detected CASR readout is employed \cite{glenn_2018}. The CASR sequence (Fig.\ \ref{figcasr}) consists of a train of phase-sensitive dynamical decoupling (DD) measurements performed by the NV sensor, with the DD microwave pulse spacing tuned to the target NMR frequency via the relation
\begin{equation}
\tau_\text{NV}=\frac{1}{4f_\text{NMR}},
\end{equation}
where we define $\tau_\text{NV}$ as the time between the $90^{\circ}$ and $180^{\circ}$ pulse ($2\tau_\text{NV}$ between $180^{\circ}$ pulses). Each DD measurement in the train is analogous to a single time point of a conventional (inductive) NMR signal acquisition.

For each DD measurement, two DD sequences are played out, referred to as ``signal'' and ``reference.'' First, the NV electronic spin state is optically initialized with a pulse of green light; an XY16-2 sequence is applied via microwave pulses; and NV PL is read out during a second green pulse (which also initializes the NV spins for the next XY16-2 sequence). The DD sequence is repeated a second time with the final $90^{\circ}$ pulse inverted, thereby detecting the opposite NV spin projection. The two readouts are subtracted and normalized to produce the final DD measurement for a given time point. This process is repeated for a series of time points to create a time course measuring the nuclear spin precession. Due to the phase-sensitive nature of CASR detection, the time course maps out a sinusoid with a beat frequency determined by the dwell time and the NMR frequency.

Inductively detected NMR often sets the detector frequency at the nuclear spin Larmor frequency, placing the spectral peak at 0 Hz. However, with CASR we typically choose a dwell time that places the detected spectral peak away from 0 Hz. For our experimental demonstration, the dwell time is between 48 and 52 $\mu$s, giving a detection bandwidth of $\sim$20 kHz and an NMR signal at $\sim$6 kHz. This choice also creates oscillatory spin-echo signals that cannot be quantified by direct integration in the time domain. We instead Fourier transform each echo and integrate the peak in the spectral domain.

\subsection{NMR Pulse Sequences}

We implement standard NMR pulse sequences such as $90^\circ$-acquire and spin-echo, but with CASR acquisition by the NV center replacing inductive acquisition by conventional pickup coils. We therefore refer to ``free-decay'' of the NMR signal, as there is no inductive detection. We perform four-step phase cycling of the transmit pulse in order to create complex time-domain data, as described in Appendix A. For most measurements, Overhauser dynamic nuclear polarization (DNP) is performed for between 500 and 1500 ms before the initial 90$^\circ$ excitation pulse \cite{carver,abragam1957}. Overhauser pulses are applied using the same microwave loop and amplifier as for pulses to manipulate NV electronic spins.

We perform homonuclear J-coupling spectroscopy with the synchronized echo (SyncE) sequence previously described elsewhere \cite{synce}. We utilize spin-echo double resonance (SEDOR) to perform heteronuclear J-coupling spectroscopy. We also implement a CPMG-SEDOR sequence in which the simultaneous echo pulses are repeated following a single $^1$H excitation pulse \cite{speier,mandal2015,Donaldson}. This allows the J-induced dynamics to be measured in a single sequence, with the J-coupling frequency then determined by fitting the resulting sinusoid or performing the Fourier transform. See section Appendix B for a detailed description of the pulse sequences.

\subsection{NMR Simulations}
We simulate SyncE spectra and SEDOR dynamics using the SPINACH package \cite{spinach}. J-coupling and chemical shift parameters for propylene carbonate are found from the literature \cite{lokesh2014}. Starting at the methyl group, chemical shifts are $\delta =$ 1.47, 4.84, 4, and 4.52 ppm for spins A, B, C, and C', respectively. J-couplings are $J_\text{AB}=$ 6.27 Hz, $J_\text{BC}=$ 7.2 Hz, and $J_\text{BC'}=$ 7.6 Hz. 

\section{Results}
\begin{figure*}[h!]
\centering
\includegraphics[width=16cm]{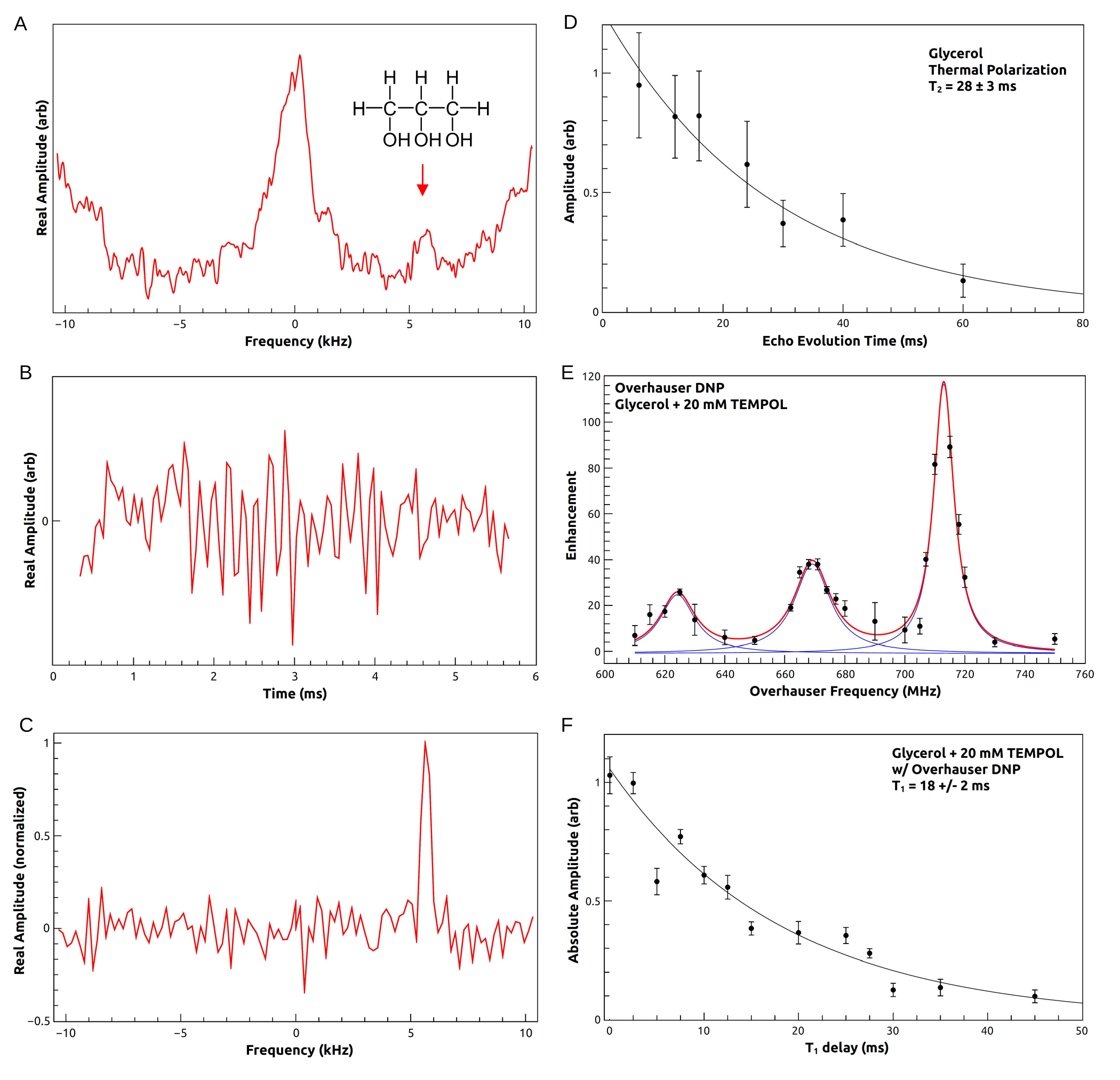}
\caption{Low-field NV-NMR of glycerol. (A) Free-decay CASR measurement of a thermally polarized sample shows a single NMR spectral line at $\sim$ 6 kHz as well as large artifacts from coil ringdown. (B) and (C) Acquisition of a spin echo and windowing of the time domain eliminates the ringdown artifact from the RF coil. (D) A series of spin-echo measurements gives T$_2$ = $28\pm3$ ms when fit to a single exponential decay. (E) Addition of 20 mM TEMPOL enables Overhauser DNP and polarization enhancements of over 100 times the thermal value. Here, the polarization enhancement increases with microwave frequency due to the gain response of the microwave amplifier. (F) T$_1$ = $18\pm2$ ms measured using a series of spin-echo measurements delayed from the polarization step, which shows the shorter relaxation time resulting from the TEMPOL radical. }\label{glycerol}
\end{figure*}

We first acquire measurements on thermally polarized glycerol in an open sample cup, with an estimated sample volume of 10 pL producing 50\% of the signal. The free-decay signal exhibits a single-sided spectrum with 820$\pm$60 Hz linewidth at a 5.68$\pm$0.01 kHz aliased resonance frequency in our spectral window (Fig.\ \ref{glycerol}A). A large artifact is also present at 0 Hz resulting from detection of the RF coil ringdown. Due to the small signal available from thermal polarization, the acquisition requires 50,000 averages, taking $\sim$2 hours. To eliminate the ringdown artifact, we switch to spin-echo measurements. Fig.\ \ref{glycerol}B and C show an example of the time-domain signal and phased spectrum, respectively, for a 3 ms echo time. The acquisition is windowed starting at 336 $\mu$s and centered at 3 ms. The linewidth is 360$\pm$60 Hz, narrower than the linewidth in the free-decay experiment, likely because we use fewer averages (15,000), and there is less magnet drift over the course of the experiment. Using a series of spin-echo measurements, we determine T$_2 = 28\pm3$ ms (Fig.\ \ref{glycerol}D), consistent with inductively detected T$_2$ measured elsewhere \cite{gach,Aroulmoji}. This NV-NMR T$_2$ implies an 11 Hz linewidth contribution from spin decoherence, with the rest of the linewidth resulting from B$_0$ inhomogeneity and drift.

To boost the NMR signal and shorten the required averaging time, we implement Overhauser DNP. Glycerol mixed with 20 mM TEMPOL shows polarization enhancements of over 100 for the strongest hyperfine peak (Fig.\ \ref{glycerol}E). Using a series of delays between the Overhauser DNP pulse and the 90$^\circ$ NMR excitation pulse, we measure longitudinal relaxation, finding T$_1 = 18\pm2$ ms (Fig.\ \ref{glycerol}F). This value is shorter than the T$_2$ found for the thermally polarized sample due to the presence of TEMPOL (T$_1$ is typically greater than T$_2$ for glycerol \cite{aweda}).

While the short T$_1$ of glycerol is convenient for establishing basic functionality, it is too short to measure J-coupling spectra, which requires longer times for J-coupling dynamics to be spectrally resolved. Turning to lower-viscosity liquids with longer T$_1$, we find that diamond heating (due to laser and microwave excitation pulses) causes the samples to quickly evaporate from the open sample cup. Hence a sealed container is used to prevent evaporation over the course of the experiment: specifically, thin-walled borosilicate or quartz tubes with 10 $\mu$m thick glass walls, originally designed for x-ray experiments, and sealed with cha-seal. The sample tube is placed against the diamond surface, providing a stable sample but with an extra 10 $\mu$m standoff distance from the NV ensemble. For a 10 $\mu$m deep NV layer, the extra standoff reduces the NMR signal strength by about 11\%, a modest loss (see Appendix E).

\begin{figure}[h!]
\centering
\includegraphics[width=9cm]{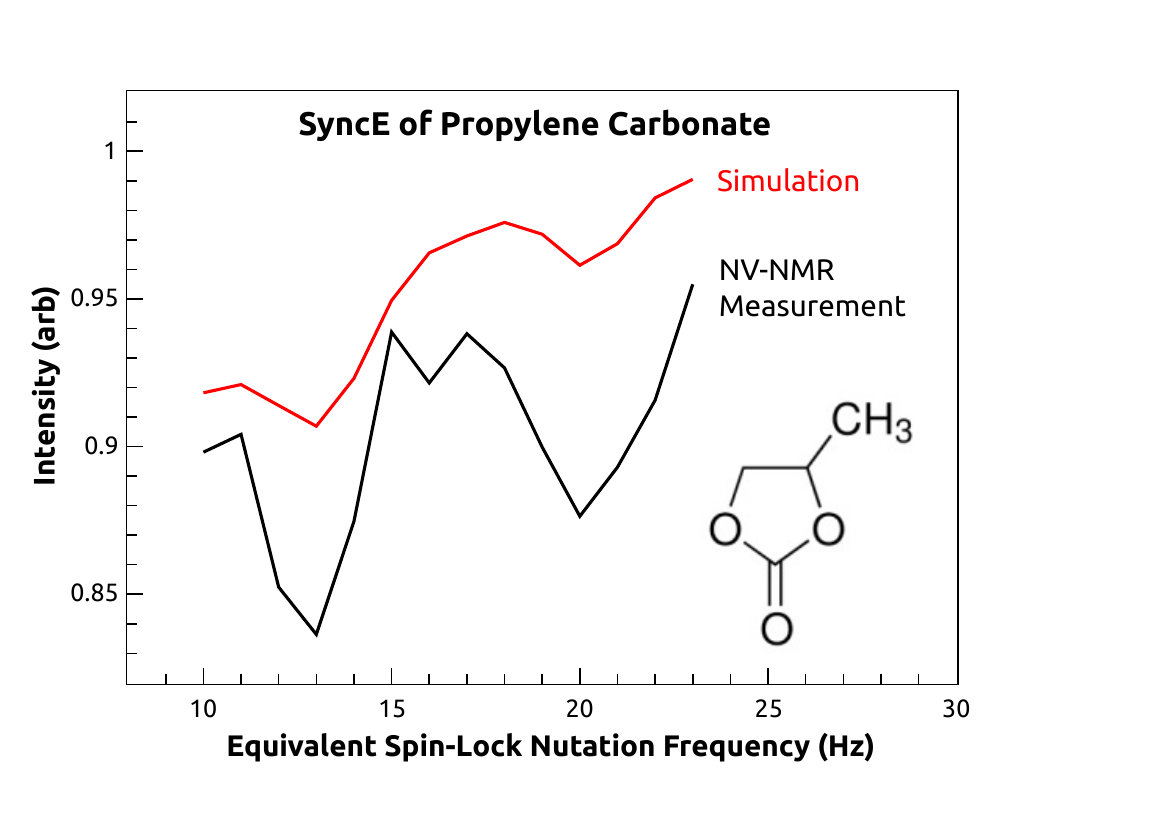}
\caption{Low-field NV-NMR of propylene carbonate measured with the Synchronized Echo (SyncE) sequence. NV-NMR amplitude dips result from the creation of NMR-dark dressed states at particular echo times (here converted to an equivalent SLIC spin-lock nutation frequency). The frequencies of the dips agree with those found from simulations using high-field J-coupling measurements. The measured dip intensity may be stronger due to additional relaxation beyond the scope of the model. }
\label{syncedata}
\end{figure}

The sealed sample tube enables us to test homonuclear J-coupling spectroscopy with SyncE on a sample of propylene carbonate containing 1.25 mM TEMPOL (Fig.\ \ref{syncedata}).  We measure two NV-NMR amplitude dips at the expected equivalent spin-lock nutation frequencies (12.8$\pm$0.2 Hz and 20.9$\pm$0.7 Hz). The dip locations match our simulation based on J-couplings determined at high fields \cite{lokesh2014}; however, the dip intensity is stronger than expected. One possible reason is stronger relaxation of the dressed states when the molecule is in close contact with the sample tube wall.

\begin{figure*}[h!]
\centering
\includegraphics[width=16cm]{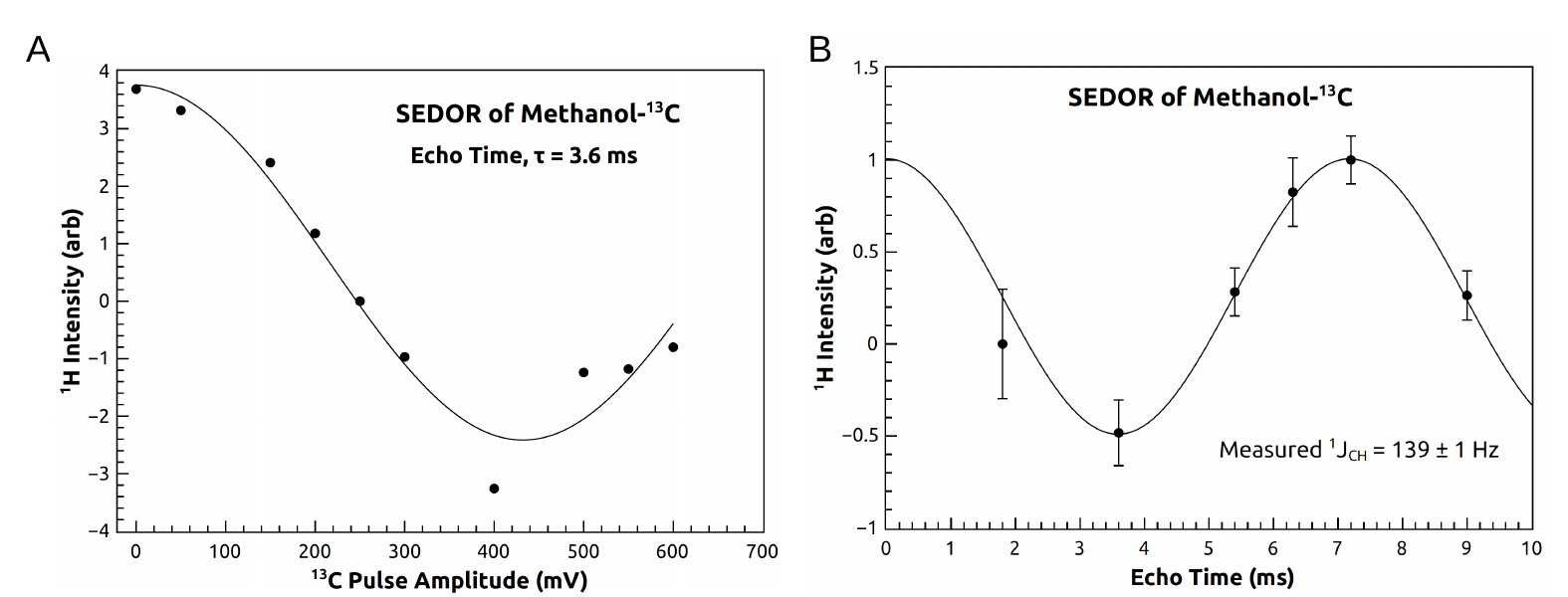}
\caption{Low-field NV-NMR measurements of methanol-$^{13}$C using spin-echo double resonance (SEDOR). (A) $^{13}$C pulse calibration at a fixed echo time. (B) SEDOR at a series of echo times, fit with the function $y=A \sin(2\pi J \tau)+y_0$ to find $^1J_\text{CH}=139\pm1$ Hz.}\label{sedor}
\end{figure*}

We next acquire NV-NMR spectra from a heteronuclear spin system, methanol-$^{13}$C, using spin-echo double resonance (SEDOR). To enable these experiments, we add a $^{13}$C coil perpendicular to the $^{1}$H coil along with a diplexer to allow use of the same AWG source and amplifier as the $^{1}$H channel. We calibrate the $^{13}$C 180$^\circ$ pulse using a spin-echo SEDOR sequence with a series of $^{13}$C pulse amplitudes. The echo time is set to $1/(2J) = 3.6$ ms, which inverts the methyl group protons when the 180$^\circ$ pulse is properly calibrated (Fig.\ \ref{sedor}A). We then acquire SEDOR measurements at a series of spin-echo evolution times to create a time-domain curve of J-coupling evolution (Fig.\ \ref{sedor}B). Fitting a sinusoid to the curve gives $^1J_\text{CH}=139\pm1$ Hz, in line with the literature value of 140.5 Hz \cite{kaseman}. 

\begin{figure*}[h!]
\centering
\includegraphics[width=16cm]{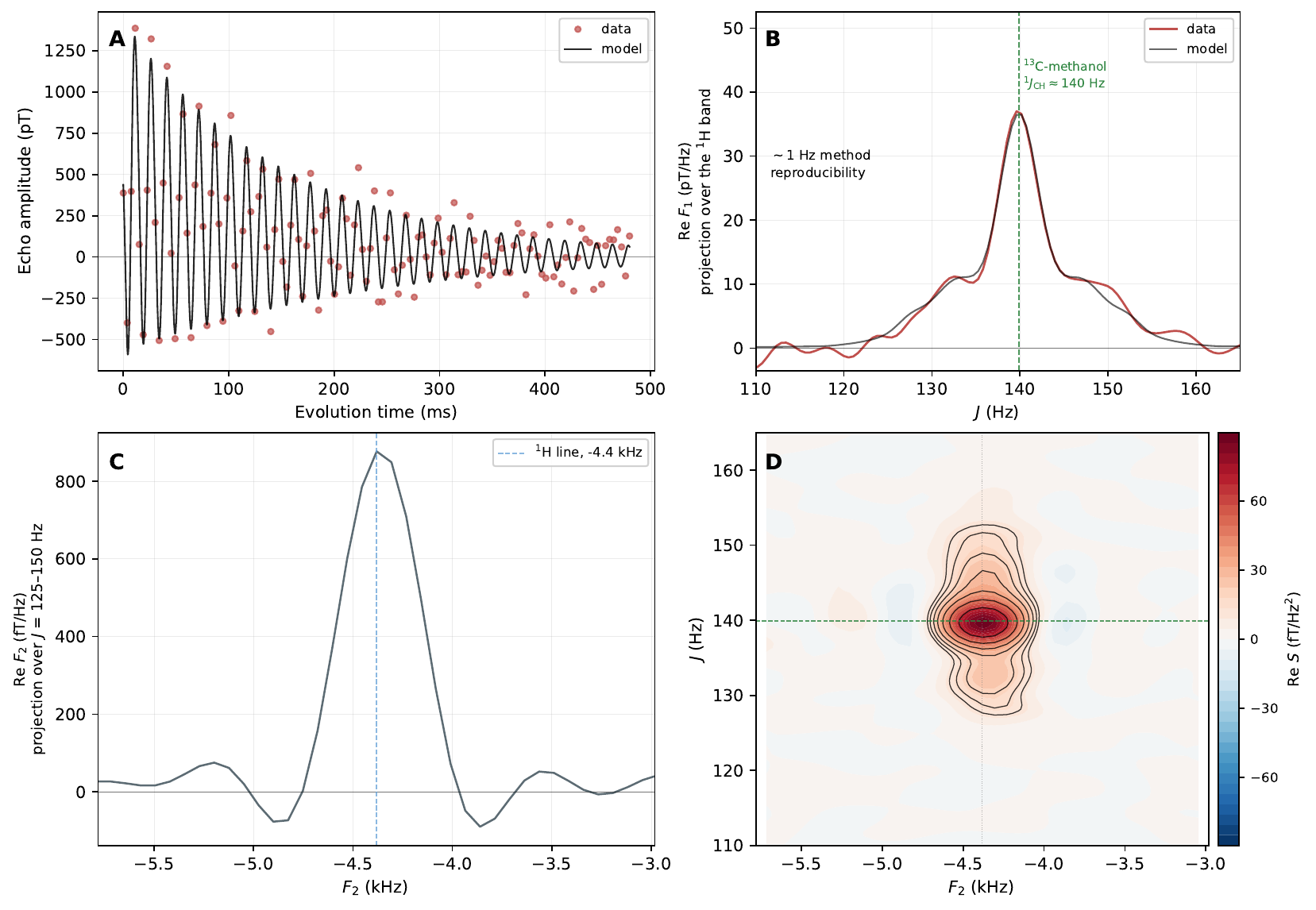}
\caption{Low-field NV-NMR measurements of methanol-$^{13}$C using CPMG-SEDOR. (A) Time-domain data from CPMG-SEDOR, fit using a single sinusoid model with exponential decay. (B) and (C) Spectral slices through the indirect and direct dimensions, respectively, showing $^1J_\text{CH}= 140\pm1$ Hz for the line at $-4362\pm5$ Hz. (F) 2D Fourier transform of the data. }\label{sedor-cpmg}
\end{figure*}

Performing a series of spin-echo SEDOR measurements is time inefficient, so we next perform CPMG-SEDOR, in which many echoes are acquired within a single pulse train. Fig.\ \ref{sedor-cpmg}A shows examples of measured echo amplitudes plotted as a function of evolution time. Oscillations are a result of J-coupling evolution on the proton spins, while the envelope decays at T$_2$ (a function of the particular CPMG echo time used). The data are analyzed with a 2D Fourier transform to produce a plot showing a clear low-noise spectrum with peaks at $-4362\pm5$ Hz in the direct dimension, representing the Larmor frequency; and at $140\pm1$ Hz in the indirect dimension, representing the J-couplings in the system. Slices through the two dimensions show linewidths of $480\pm30$ Hz in the direct dimension and $4.6\pm0.8$ Hz in the indirect dimension, with the latter limited by T$_2$(Fig.\ \ref{sedor-cpmg}B-D).

\begin{figure*}[h!]
\centering
\includegraphics[width=16cm]{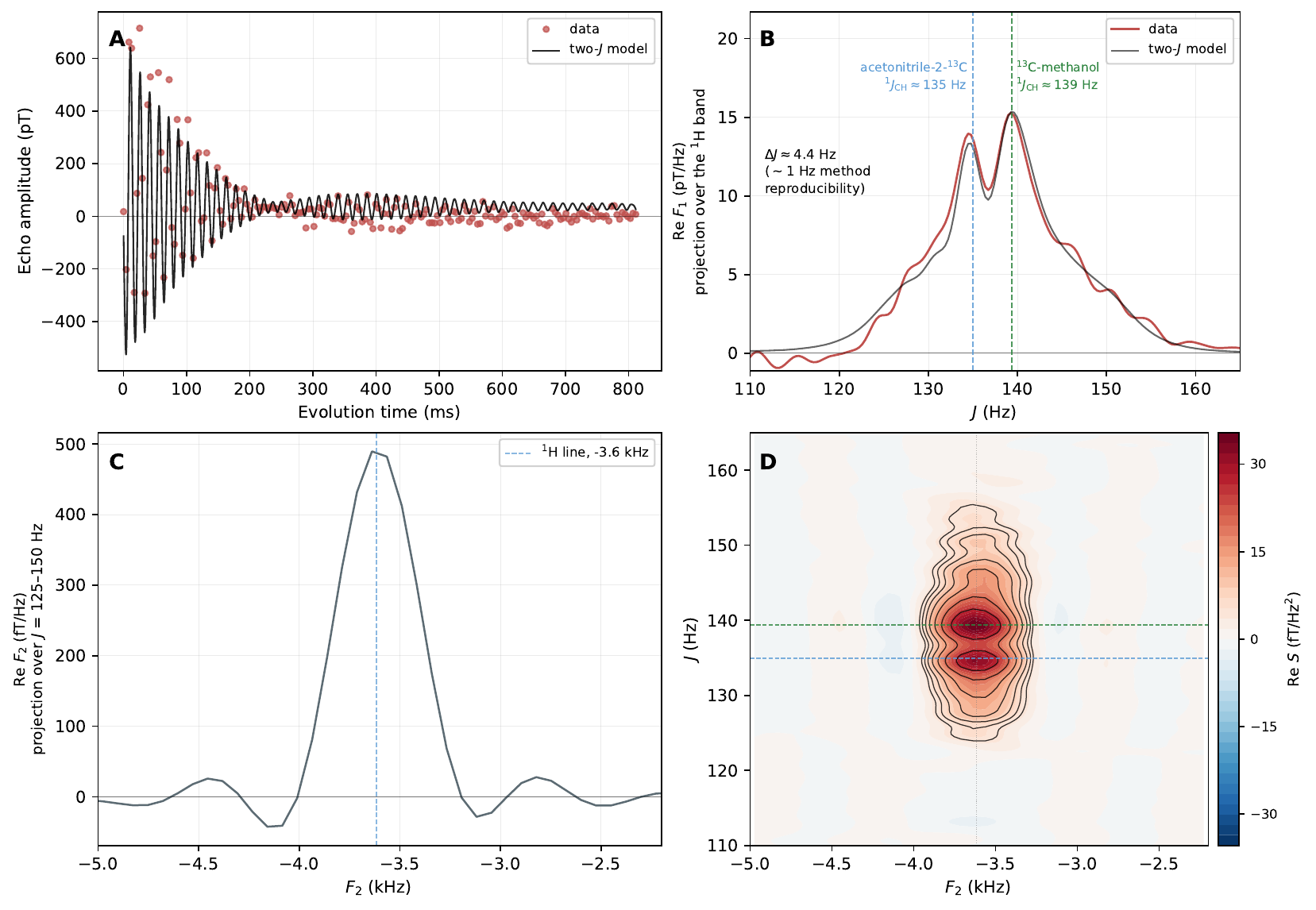}
\caption{Low-field NV-NMR measurements of a methanol-$^{13}$C/acetonitrile-2-$^{13}$C mixture using CPMG-SEDOR. (A) Time-domain data from CPMG-SEDOR, fit using the sum of two sinusoids with exponential decay. (B-C) Spectral slices through the indirect and direct dimensions, respectively, showing two peaks at $^1J_\text{CH}=135\pm1$ Hz and $^1J_\text{CH}=139\pm1$ Hz for the line at $-3608\pm3$ Hz. (F) 2D Fourier transform of the data showing two peaks.}\label{sedor2}
\end{figure*}

Finally, to show that our low-field NV-NMR system can measure two different species simultaneously, we acquire CPMG-SEDOR measurements of a mixture of methanol-$^{13}$C and acetonitrile-2-$^{13}$C (Fig.\ \ref{sedor2}). From literature, acetonitrile-2-$^{13}$C has a J-coupling $^1J_\text{CH}$ between 131 and 141 Hz for the methyl group, with variability due to solvent and temperature effects \cite{zf-nmr,comagnetometer}. In this mixture, we find $^1J_\text{CH} = 135\pm1$ Hz for acetonitrile and $139\pm1$ Hz for methanol, with the peaks distinguishable but not fully resolved.

\section{Discussion}

We have demonstrated that valuable spectral information in small ($\sim$10 pL) samples can be measured with NV-NMR even at low magnetic fields (21 mT), enabling chemical identification based on J-coupling and connectivity. More complex pulse sequences, such as INEPT, can also determine spin multiplicity, as the time-domain dynamics of polarization transfer from $^1$H to $^{13}$C depend on the number of coupled protons. However, using conventional INEPT with NV-NMR would require CASR detection of $^{13}$C, which produces a smaller signal due to carbon's lower gyromagnetic ratio. 

Low-field J-coupling spectroscopy lends itself to small-scale on-chip NMR systems using either inductive or NV detection \cite{ha2014, ariando2019, Zhou2024, Zhang2024}. The field homogeneity and stability requirements are greatly reduced compared with conventional NMR: e.g., our NV-NMR demonstration does not include shim coils or bias magnet stabilization, which add significantly to system complexity. High-field NMR requires sub-ppm field homogeneity and stability to resolve small chemical shifts and J-coupling features on the order of Hz in a $\sim$60 MHz (>1.5 T) or higher bias field. For J-coupling spectroscopy, the homogeneity requirement is instead determined by the effect of diffusion-induced T$_2$ relaxation during the spin-echo or CPMG sequence \cite{kleinberg,ronczka}, as T$_2$ must remain sufficiently long for J-induced evolution to play out. Additionally, slow bias-field drifts during averaging have minimal effect and can be corrected via occasional interleaved calibration measurements rather than real-time magnet stabilization. 

Resolution in the J-coupling dimension is also limited by the intrinsic T$_2$ of the sample, which is determined by temperature, viscosity, and molecular structure, as well as by TEMPOL concentration (used for Overhauser DNP of the sample). Increasing TEMPOL concentration produces more Overhauser enhancement but also speeds up relaxation, lowering J-coupling resolution. Optimization is required to find a balance between the two needs. Accuracy is also affected by pulse errors, particularly for CPMG-SEDOR where they can accumulate for a large number of pulses. We discuss the effects of pulse errors in Appendix C.

Other J-coupling spectroscopy techniques can also be used for NV-NMR, each with their own strengths and weaknesses. For example, Smits \textit{et al.}\ demonstrated two-dimensional correlation spectroscopy (COSY) NV-NMR of 1,4-difluorobenzene \cite{Smits2019}. This provides similar J-coupling information to SEDOR, but requires high field homogeneity, as it is based on the free-decay spectrum rather than spin echoes. Omar \textit{et al.}\ recently performed zero- to ultralow-field (ZULF) spectroscopy at nearly zero field using an NV sensor rather than the optically pumped magnetometer that is typically utilized \cite{Omar2026,Theis2013}. While very high resolution J-coupling spectra can be achieved with ZULF, it is only sensitive to heteronuclear spin systems, requiring enriched samples or high spin polarizations to be sensitive to natural isotopomers \cite{Andrews2025}.

Finally, our low-field NV-NMR demonstration suggests that single-cell metabolic imaging should be possible, in principle. Recent inductive NMR work has demonstrated measurements of pyruvate to lactate interconversion at low magnetic fields using J-resolved sequences in the absence of chemical shift resolution \cite{Korchak2021, Rodriguez2025}. Implementing J-coupling spectroscopy combined with optical real-space imaging of NMR signals using a wide-field NV-NMR microscope \cite{DeVience2015} would enable the metabolism of individual cells or tissues to be mapped and correlated with other imaging techniques such as targeted fluorescence. This unique capability will likely require integration of J-coupling NV-NMR spectroscopy, SABRE hyperpolarization, microfluidic delivery, quantum-logic enhanced sensing, and NV widefield NMR imaging, all of which have now been demonstrated individually with NV centers \cite{DeVience2015, sabre_nv, Briegel2025, Smits2019, Yin, Droid}. 

\section{Acknowledgments}
We acknowledge support for this work from the Gordon and Betty Moore Foundation, the NASA NSTGRO graduate fellowship program, and the University of Maryland Quantum Technology Center.

\appendix
\renewcommand{\thefigure}{B.\arabic{figure}}
\setcounter{figure}{0}      
\section{Creation of One-Sided Spectra}
NMR spectrometers typically use heterodyne detection of the induced magnetic signal, which produces complex-valued data that are converted to a complex spectrum via the Fourier transform. Compared with purely real data, this has two advantages: (1) The Nyquist bandwidth is given by the full sampling rate $F_s$ instead of $F_s/2$; and (2) out-of-band noise is not folded into the window, so SNR is increased by $\sqrt{2}$. However, NV-NMR using CASR \cite{glenn_2018} only produces real-valued time-domain data, and previous demonstrations with CASR have typically presented spectra in magnitude mode from just one side of the spectral window (as the other side is a mirror-image). 

To create complex-valued time-domain data and utilize the full spectral window, we perform a simple phase cycling technique. We take four acquisitions cycling through four different transmit phases ($\theta = 0^\circ, 90^\circ, 180^\circ, 270^\circ$). Since all acquisitions are identical and have the same detection phase, we then assign each of the four acquisitions the appropriate coefficient $C = \exp(i \theta)$ to make the data complex. This gives
\begin{equation}
  S = \sum_{j}C(\theta_j)S_j=S_1+iS_2-S_3-iS_4  .
\end{equation}
Using two steps ($\theta = 0^\circ, 90^\circ$) is also possible but less robust to imperfections.

Because real and imaginary parts are acquired serially rather than in parallel, the $\sqrt{2}$ SNR improvement of advantage (2) is negated by the two-fold longer experiment time. With our technique, two real-valued NV-NMR CASR measurements produce the same SNR as a single four-phase complex-valued measurement but with a spectral window of half the bandwidth.

\section{Pulse Sequences for J-Coupling Spectroscopy}

\begin{figure*}[h!]
\centering
\includegraphics[width=16cm]{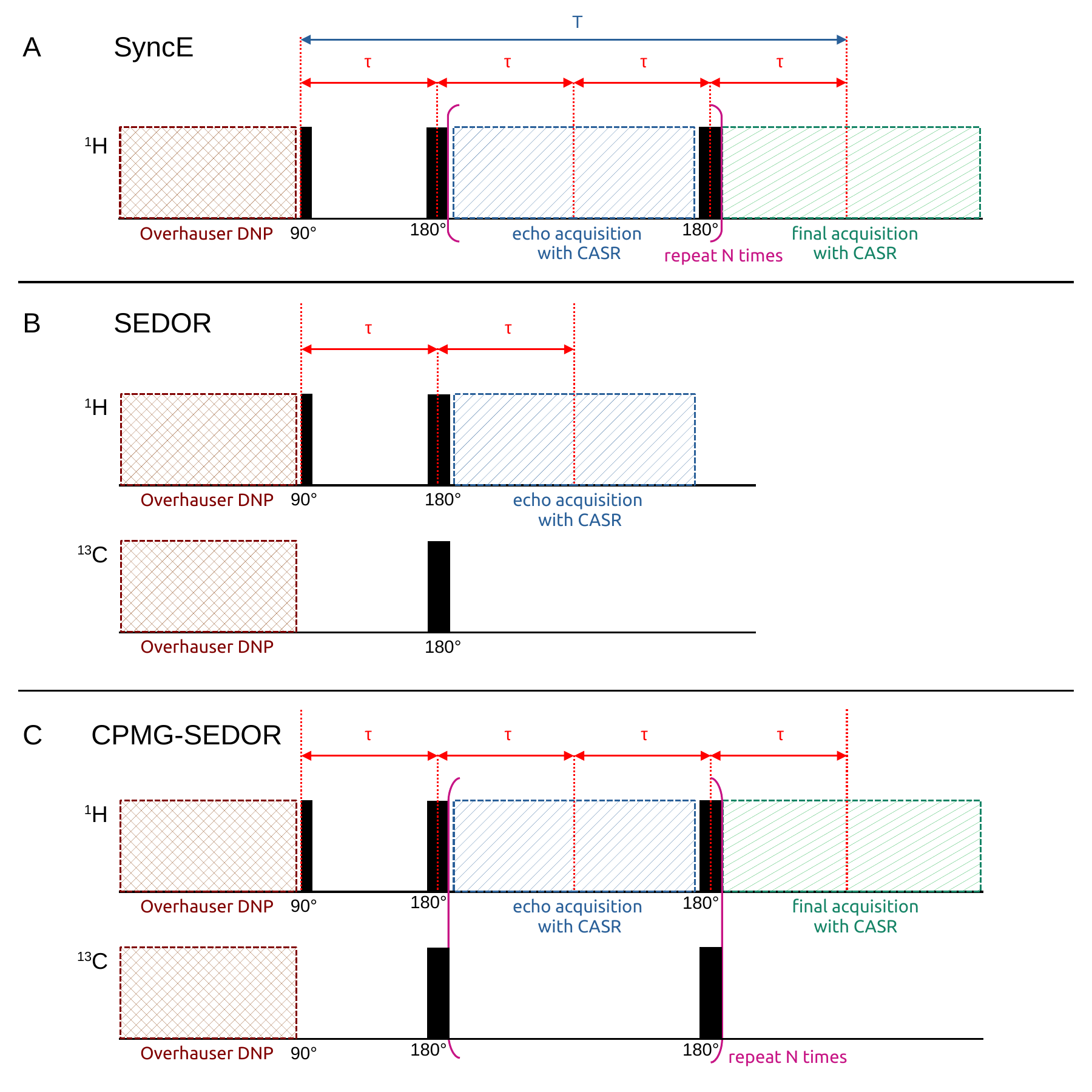}
\caption{NV-NMR pulse sequences for J-coupling spectroscopy at low field. In all cases, we first apply Overhauser DNP. (A) For synchronized echo (SyncE), we perform a series of CPMG sequences, each with its own echo time $\tau$, but keeping the total time T=2$\tau$(N+1) constant. Echoes are acquired within the sequence and can be averaged to increase SNR or used separately to create a 2D spectrum. (B) For spin-echo double resonance, we perform a series of single spin-echo measurements with different $\tau$, but now an inversion pulse is also applied to the heteronucleus. (C) CPMG-SEDOR is a multi-pulse version of SEDOR in which a series of inversions are performed and echoes are acquired within the sequence.}\label{figpulses}
\end{figure*}

Synchronized Echo (SyncE) is a multi-pulse adaptation of the spin-lock induced crossing (SLIC) method \cite{slic1, slic2}, which detects transitions between dressed-state energy levels in the weak-coupling regime ($J\gg\Delta f$, i.e.\ J-coupling much larger than chemical shift differences). For SLIC, a series of measurements are taken at increasing spin-lock nutation frequencies $\nu_\text{nut}$, and dips in amplitude occur in the signal when $\nu_\text{nut}$ matches these transition frequencies. SyncE instead acquires a series of CPMG measurements, with decreasing echo time and increasing echo number to keep the total evolution time constant (Fig.\ \ref{figpulses}A). Echoes are averaged to increase the SNR. SyncE and SLIC spectra are nearly identical, with the equivalent spin-lock nutation frequency given by $\nu_{n}=1/(4\tau_n)$, where $\tau_n$ is the echo time for the $n^\text{th}$ acquisition in the series. 

We utilize spin-echo double resonance (SEDOR) to perform heteronuclear J-coupling NV-NMR spectroscopy at low field. SEDOR measures the remaining $^1$H signal following a simultaneous echo pulse on $^1$H and the heteronucleus, in our case $^{13}$C. This is similar to INEPT, but with both excitation and detection on $^1$H (Fig.\ \ref{figpulses}B). Unlike INEPT, the $^1$H multiplicity does not affect the signal dynamics, and a plot of intensity vs.\ echo time is a single sinusoid at the J-coupling frequency \cite{molin1987}. We also implemented a CPMG-SEDOR sequence in which the simultaneous echo pulses are repeated following a single $^1$H excitation pulse (Fig.\ \ref{figpulses}C) \cite{speier,mandal2015,Donaldson}. This allows the J-induced dynamics to be measured in a single sequence, with the J-coupling frequency then determined by fitting the resulting sinusoid or performing the Fourier transform.

\section{Effects of Pulse Errors}

We find that CPMG-SEDOR is sensitive to $^1\text{H}$ pulse errors. In a series of experiments with echo times $\tau = 1.86$ ms, $\tau = 2.43$ ms and $\tau = 3.26$ ms, we find $J = 148.4$ Hz, $J = 136.9$ Hz, and $J = 136.8$ Hz, respectively (Fig.\ \ref{pulse_error}A). The average of the three measurements is $140.7\pm6.6$ Hz, in agreement with the literature value, albeit with a high uncertainty. Fitting the $\tau = 1.86$ ms data in the time domain also confirms our measured value. However, the measured value of J should be independent of the echo time used; thus we investigated whether measurement errors are caused by pulse errors in the CPMG train.  Via simulation, we find that for $\tau = 1.86$ ms, the measured J-coupling value increases as pulse error gets larger, while for $\tau = 2.43$ ms and $\tau = 3.26$ ms, the measured J-coupling decreases. A pulse error of $\sim$10-15\% gives the values we measure for the three echo times in this experiment. Simulations show that an echo time $\tau = 2.01$ ms should give minimal measurement error for pulses of up to at least 40\% error.

\renewcommand{\thefigure}{C.\arabic{figure}}
\setcounter{figure}{0}     
\begin{figure}[h!]
\centering
\includegraphics[width=8cm]{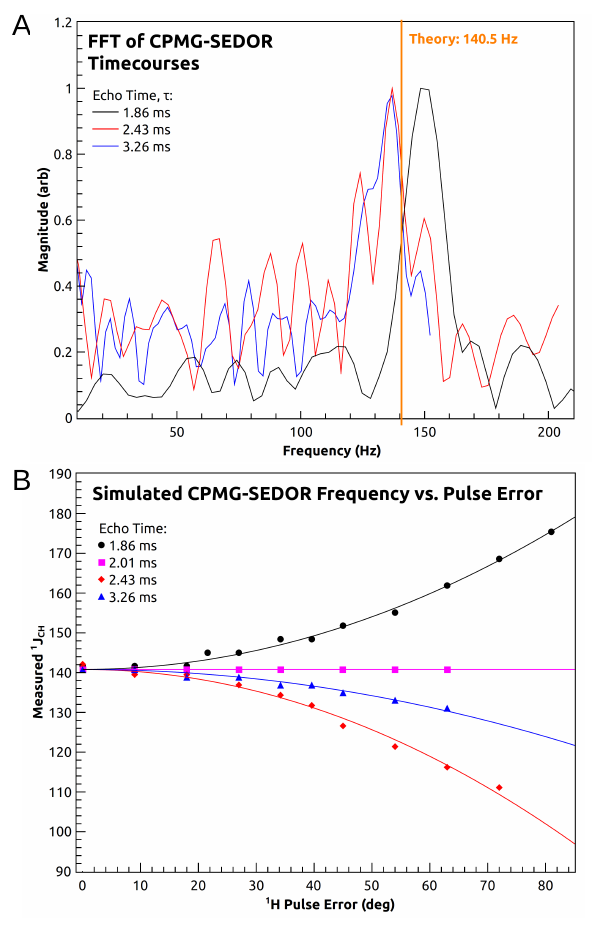}
\caption{Effects of proton pulse errors on low-field NV-NMR CPMG-SEDOR. (A) Measurements with different echo times produce peaks at slightly different frequencies centered around the correct frequency. (B) Simulations show that the most likely cause is $^1$H pulse errors, which can have a strong effect on the measured frequency. However, the effect can be minimized by choosing an appropriate echo time.}\label{pulse_error}
\end{figure}

\section{Processing of SyncE Data}
We acquire SyncE J-coupling data using a series of CPMG trains with different echo times but constant total evolution time. For each CPMG train $n$, the echo time is $\tau_n$ and the corresponding equivalent spin-lock nutation frequency $\nu_{n}$ is given by $\nu_{n}=1/(4\tau_n)$. Within each train $n$, every echo is windowed with 104 points and converted to the spectral domain via the fast Fourier transform. The spectra are phased based on the phase of the first echo. Three points of the spectral peak are integrated and the integrals $I(k)$ of each echo $k$ are corrected for T$_2$ according to

\begin{equation}
I(k)_\text{corr}=I(k) \exp[2(k-1)\tau_n/T_2]
\end{equation}
where $I(k)_\text{corr}$ is the corrected echo integral and $T_2$ is the relaxation time. We use $T_2$=500 ms. The spectral intensity $S$ is found by summing $I$ over all $k$. This is performed for all the CPMG trains to create a spectrum $S(\nu_n)$. 

However, because T$_2$ is a function of echo time and is longer at shorter echo times, the spectral baseline must be corrected. We use a function of the form 

\begin{equation}
S(\nu_n)_\text{corr}=\frac{S(\nu_n)}{A+B\nu_n} 
\end{equation}
where $S(\nu_n)_{corr}$ is the corrected spectrum and A and B are empirical coefficients. We used A=0.75 and B=0.015.

\section{Sample Signal Estimation for Sensor Geometry}
Using a sample tube rather than placing the sample directly on the NV-diamond surface is expected to decrease the total NMR signal at the NV layer. However, we find that for thin-walled tubes the effect is modest. Using a numerical calculation based on Ref.\ \cite{glenn_2018}, we estimate the signal strength at the NVs for thermally polarized water on the surface and in our sample tube. The dipole-dipole coupling at an NV is given by 

\begin{equation}
\vec{B}\cdot\hat{\mu}_\text{NV} =C \int \frac{1}{r^3}[3(\hat{r}\cdot\hat{m}(t))(\hat{r}\cdot\hat{\mu}_\text{NV})-\hat{m}(t)\cdot\hat{\mu}_\text{NV}]dV ,
\end{equation}
where C is a prefactor given by
\begin{equation}
C=\frac{\mu_0 \hbar \gamma_p}{4\pi}\rho_p\left[1-\exp(-\frac{\hbar\gamma_p B_0}{k_BT})\right].
\end{equation}
Here $\rho_p$ is the density of protons, 111 M for water, $k_B$ is the Boltzmann constant, $B_0$ is the static magnetic field, $\gamma_p$ is the gyromagnetic ratio of the proton, and $T$ is the temperature.

To numerically integrate, we create a 3D grid of (1 $\mu$m)$^3$ elements. A row of ten NV cubes represents a column of NVs 10 $\mu$m deep, reflecting a cross section of the NV layer. Sample blocks containing water are then placed at the appropriate distances adjacent to the diamond, either starting at the diamond surface (no tube) or beyond the tube wall. The volume is integrated to a distance of up to 500 $\mu$m above the diamond surface and $\pm$ 250 $\mu m$ in x and z. We find that the NMR signal amplitude detected by an average NV is about 35 pT for the sample directly on the surface and 31 pT for the sample in a tube. 

We also investigate the NMR signal for different size cube-shaped integration volumes above the diamond surface and find that somewhat more sample volume is needed to produce the maximum signal when the sample is in the tube (Fig.\ref{size}). When the sample is on the surface, 95\% of the maximum signal is attained for a (130 $\mu$m)$^3 = 2$ nL volume, whereas (200 $\mu$m)$^3 = 8$ nL is needed for the sample in the tube. However, due to the non-linear relationship, 50\% of the signal comes from a much smaller volume, about 10 pL for sample on the surface and 50 pL for sample in a tube, consistent with Ref.\ \cite{glenn_2018}. 

\renewcommand{\thefigure}{E.\arabic{figure}}
\setcounter{figure}{0}     
\begin{figure*}[h!]
\centering
\includegraphics[width=16cm]{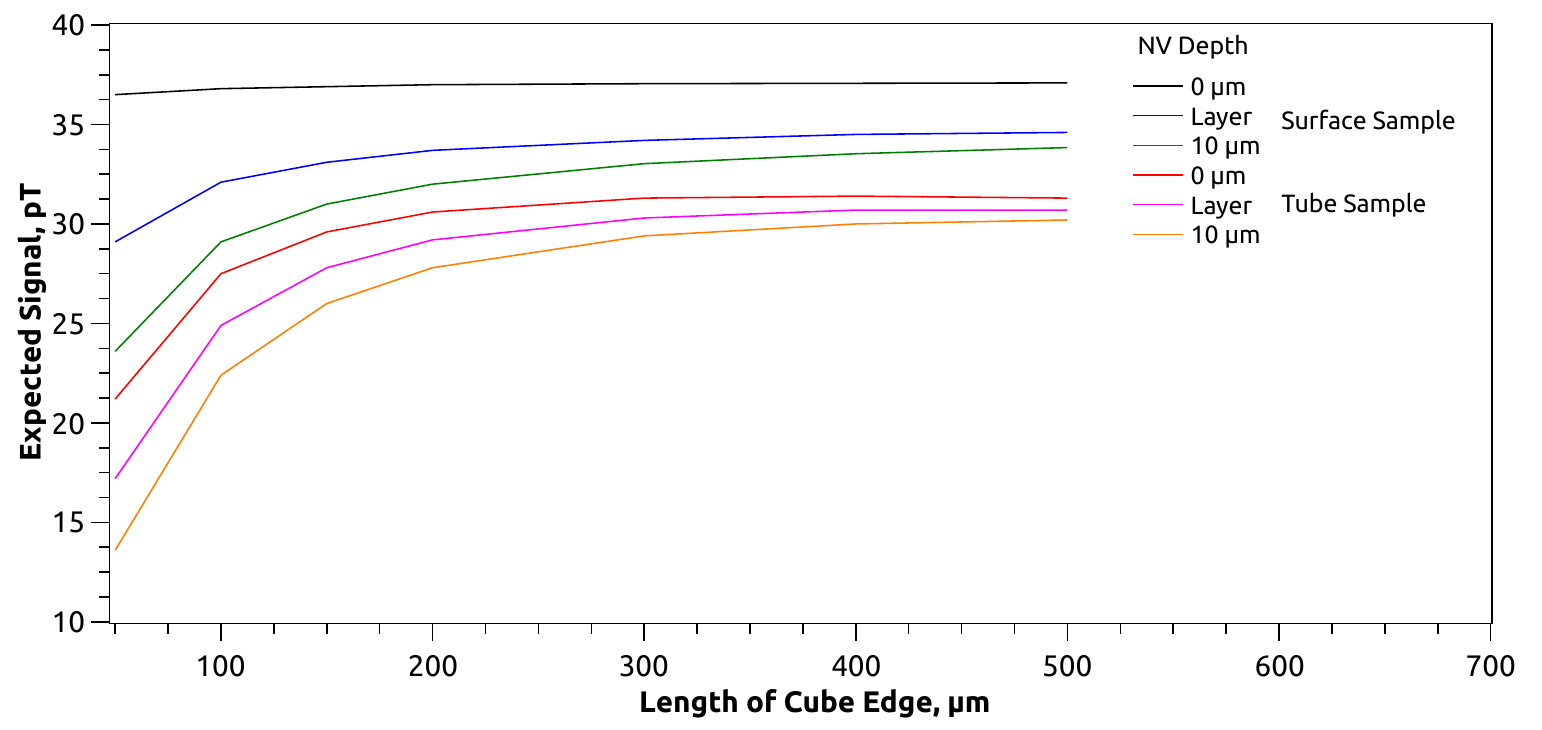}
\caption{Low-field NV-NMR signal vs.\ sample size for water directly on the NV-diamond surface and in a tube with 10 $\mu$m thick walls. As the size of the sample increases, the signal plateaus, indicating that most of the signal is coming from spins near the diamond. However, maximum signal can be attained at smaller volumes when the sample is directly on the surface and when NVs are closer to the surface. For sufficiently large samples, the effect of the tube is modest, with signals only 11\% smaller then those from a sample on the surface. Calculations are performed for NVs at the diamond surface, 10 $\mu$m deep, and averaged for a 10 $\mu$m NV layer.}\label{size}
\end{figure*}

\bibliographystyle{elsarticle-num-names} 
\bibliography{nv-nmr-refs}

\end{document}